\documentclass[twocolumn,showpacs,amsmath,amssymb,prc]{revtex4}
\usepackage{mathrsfs}

\usepackage{graphicx,color}
\usepackage{dcolumn}
\usepackage{bm}
\usepackage{CJK}

\begin{document}

\title{Constraint on the cosmic age from the solar $r$-process abundances}

\author{T. H. Heng$^1$}
\author{X. D. Xu$^2$}
\author{Z. M. Niu$^1$}\email{zmniu@ahu.edu.cn}
\author{B. H. Sun$^2$}
\author{J. Y. Guo$^1$}

\affiliation{$^1$School of Physics and Material Science, Anhui University,
             Hefei 230039, China}
\affiliation{$^2$School of Physics and Nuclear Energy Engineering, Beihang University,
             Beijing 100191, China}
\date{\today}

\begin{abstract}
The cosmic age is an important physical quantity in cosmology. Based on the radiometric method,
a reliable lower limit of the cosmic age is derived to be $15.68\pm 1.95$ Gyr by using the
$r$-process abundances inferred for the solar system and observations in metal-poor stars. This
value is larger than the latest cosmic age $13.813\pm 0.058$ Gyr from Planck 2013 results, while
they still agree with each other within the uncertainties. The uncertainty of $1.95$ Gyr mainly
originates from the error on thorium abundance observed in metal-poor star CS 22892-052, so
future high-precision abundance observations on CS 22892-052 are needed to understand this age
deviation.
\end{abstract}

 \maketitle
The cosmic age is a typical parameter in cosmology. According to the Big Bang cosmology, the
cosmic age usually refers to the time elapsed since the Big Bang itself. Based on the cosmic
microwave background (CMB) temperature spectra, the cosmic ages are derived to be $13.772\pm
0.059$ Gyr~\cite{Bennett2013ApJSS} based on the nine-year Wilkinson microwave anisotropy probe
(WMAP) observations and $13.813\pm 0.058$ Gyr based on the Planck
measurements~\cite{Planck2013arXiv}. Besides the cosmological method, there are several
independent methods, such as the radiometric method~\cite{Cowan1999ApJ, Goriely2001AA,
Cayrel2001Nature, Wanajo2002ApJ} and the stellar evolution method~\cite{Chaboyer1996Science,
Gratton1997ApJ, Krauss2003Science, Hansen2002ApJ, Hansen2004ApJSS}, which can be used to determine
the cosmic age as well. These age estimates set a lower limit on the cosmic age and hence can
serve as an independent check for the age derived from cosmological model.

The radiometric method is independent of the uncertainties associated with fluctuations in the
microwave background~\cite{Bennett2013ApJSS} or models of stellar
evolution~\cite{Krauss2003Science, Hansen2004ApJSS}. In this method, the age is determined by
comparing the current abundances of radioactive nuclei with the initial abundances at their
productions. This method can be traced back to the early twentieth century, when Rutherford
outlined the essential features of this method~\cite{Rutherford1929Nature}. For determining the
cosmic age, the lifetimes of the radioactive nuclei should be the order of the cosmic age, such as
the long lived radioactive nuclei $^{232}$Th and $^{238}$U (For brevity, we will use Th and U to
represent $^{232}$Th and $^{238}$U hereafter). It is known that the Th and U are synthesized by
the astrophysical rapid neutron-capture process ($r$-process)~\cite{Burbidge1957RMP,
Cameron1957CRR}. Therefore, the initial Th and U abundances can be predicted with the $r$-process
simulations. Actually, large efforts have been made on the calculation of the production rates of
Th and U as a function of time over the galactic evolution. Based on simple description for the
history of galactic nucleosynthesis, the uranium and thorium (U/Th) chronometer was used to deduce
the cosmic age~\cite{Fowler1960AP, Cameron1962Icarus}.

The abundances in metal-poor halo stars are usually not influenced by the Galactic chemical
evolution, so the radioactive dating technique based on the metal-poor halo stars can be used as a
relatively reliable dating technique for the Universe. The radioactive element Th was detected in
the $r$-process enhanced metal-poor halo star CS 22892-052 for the first
time~\cite{Sneden1996ApJ}, and it was also observed in many other metal-poor stars, e.g.,  HD
115444~\cite{Westin2000ApJ} and HD 221170~\cite{Ivans2006ApJ}. For the element U, it was firstly
detected in the CS 31802-001~\cite{Cayrel2001Nature}. However, due to the weakness of U lines and
severe blending issues, so far U was only observed in two other metal-poor stars, namely, BD
+17$^\circ$3248~\cite{Cowan2002ApJ} and HE 1523-0901~\cite{Frebel2007ApJ} except CS 31082-001.
With these abundance observations, the ages of these metal-poor stars can be estimated from Th/X,
U/X, or Th/U chronometers (X represents a stable element). Since the very metal-poor stars were
usually formed at the early epoch of the Universe, their ages can serve as a lower limit of the
cosmic age.

The radiometric method can avoid the uncertainties of Galactic chemical evolution model when it is
applied to the metal-poor halo star. However, because the $r$-process site is still in debate and
large mounts of neutron-rich isotopes which involved in $r$-process are still out of the reach of
experiments, the initial abundances which are obtained from theoretical $r$-process calculations
have large uncertainties. These lead to large uncertainties in the age
estimates~\cite{Cowan1999ApJ, Goriely2001AA, Zhang2012APS}. A way to avoid the uncertainties in
the theoretical $r$-process calculations is to employ solar $r$-process abundances at the time
when the Solar System became a closed system to approximate the initial $r$-process abundances,
since it is found that the solar $r$-process abundances at that time is close to the initial
$r$-process elemental abundances~\cite{Niu2009PRC}.

By subtracting the solar $s$-process abundances from the observed total solar abundances, the
solar $r$-process abundances have been well determined not only based on the classical
approach~\cite{Sneden1996ApJ, Burris2000ApJ, Simmerer2004ApJ} but also based on the more
sophisticated $s$-process nucleosynthesis model in low-mass asymptotic giant branch (AGB)
stars~\cite{Arlandini1999ApJ}. Moreover, the accurate abundances of rare earth (RE) elements in
five metal-poor stars CS 22829-052, CS 31082-001, HD 115444, HD 221170, and BD +17$^\circ$3248 are
derived recently~\cite{Sneden2009ApJ}. With these new data and the solar $r$-process abundances,
one can get a reliable lower limit for the age estimate, which is independent on the $r$-process
model and the theoretical models in nuclear physics, and hence can set a strict constraint on the
lower limit of the cosmic age.

In this paper, we first give a brief introduction to the basic formulas for age estimate and the
corresponding error treatment. Then the consistency between the observations from metal-poor halo
stars and the solar $r$-process abundances is carefully checked in a novel view. Furthermore, the
age estimates are made by taking the solar $r$-process abundances to approximate the initial
$r$-process abundances and compared with the cosmic age determined with the CMB observations.

For the metal-poor halo stars, the radiometric method is independent on the Galactic chemical
evolution. Then the time evolution of the abundance of a radioactive element $i$ follows the
exponential decay, i.e.
\begin{equation}\label{Eq:ExponDecay}
  N_i(t) = N_0 e^{-\lambda_i t},
\end{equation}
where the $N_i, N_0$, and $\lambda_i$ are the abundance observed at present, the initial abundance at its production, and the decay constant of element $i$. According to Eq.~(\ref{Eq:ExponDecay}), the time elapsed $t$ since the production of the radioactive element can be determined from:
\begin{eqnarray}\label{Eq:Age}
  t &=& 46.7 [\log_{10}(\textrm{Th/X})_{0}-\log_{10}(\textrm{Th/X})_{\textrm{obs}}]~\textrm{Gyr}, 
\end{eqnarray}
where ($\textrm{Th/X})_{0}$ and $(\textrm{Th/X})_{\textrm{obs}}$ denote the initial abundance ratio and the observed abundance ratio at present. From Eq.~(\ref{Eq:Age}), it is clear that the uncertainty of $t$ originates from the observation error and the error in the initial abundance, i.e.
\begin{equation}\label{Eq:Err}
   \delta t
  = 46.7 \sqrt{ [\delta \log_{10}(\textrm{Th/X})_{0}]^{2}
               +[\delta \log_{10}(\textrm{Th/X})_{\textrm{obs}}]^{2} }
\end{equation}
with
\begin{equation}
   \delta \log_{10}(\textrm{Th/X})
  =\sqrt{(\delta \log_{10}\textrm{Th})^2+(\delta \log_{10}\textrm{X})^2},
\end{equation}
where $\delta A$ denotes the error corresponding to the physical quality $A$.

In this work, the initial abundances are approximated by the solar $r$-process abundances which are taken from Ref.~\cite{Simmerer2004ApJ}, while the abundances of Th is updated with the data in Ref.~\cite{Lodders2003ApJ}. For the present observable abundance ratios $(\textrm{Th/X})_{\textrm{obs}}$, the average abundances of rare earth elements in five very metal-poor halo stars CS 22892-052, CS 31082-001, HD 115444, HD 221170, and BD +17$^\circ$3248 are employed~\cite{Sneden2009ApJ}, while the abundance of Th is taken from meta-poor star CS 22892-052~\cite{Sneden2003ApJ}. The corresponding errors of the average abundances of the rare earth (RE) elements is evaluated by $\delta \log_{10}\textrm{X}=\sqrt{\sum_{i=1}^5\sigma_{i}^{2}}/5$, where $\sigma_{i}$ is the abundance error in the five metal-poor stars.

\begin{figure}[h]
  \includegraphics[width=8cm]{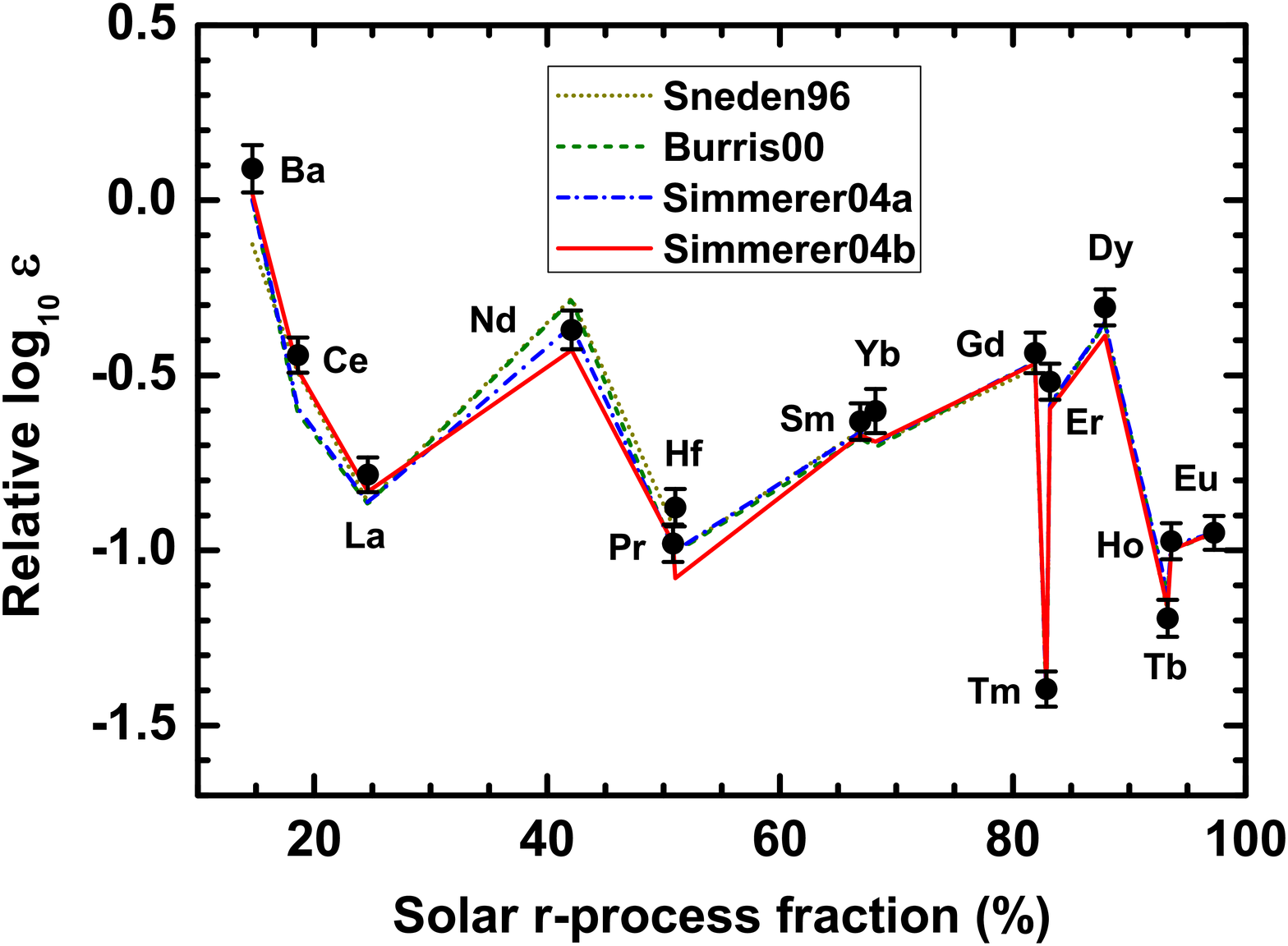}
  \caption{(Color online) Comparison of the solar $r$-process abundances and the elemental abundances obtained from metal-poor halo stars CS 22892-052, CS 31082-001, HD 115444, HD 221170, and BD +17$^\circ$3248. The filled circles represent the average abundances of five metal-poor halo stars. By subtracting the $s$-process abundances predicted by the classical $s$-process approach, the corresponding $r$-process abundances (scaled to Eu data) in Refs.~\cite{Sneden1996ApJ, Burris2000ApJ, Simmerer2004ApJ} are shown by the dotted, dashed, and dash-dotted lines, respectively. The solid line denotes the solar $r$-process abundances (scaled to Eu data) obtained with the stellar $s$-process model in Ref.~\cite{Arlandini1999ApJ} while updated by the data in Ref.~\cite{Simmerer2004ApJ}.}\label{fig1}
\end{figure}
By comparing the solar $r$-process abundance distribution and those from metal-poor halo stars, a similar abundance distribution has been found for the elements above Ba ($Z=56$)~\cite{Sneden1996ApJ, Sneden2008ARAA}. Since the solar $r$-process abundances are usually obtained by subtracting the calculated solar $s$-process abundances from the observed total solar abundance, some uncertainties are inevitable for the solar $r$-process abundances. In Fig.~\ref{fig1}, four sets of solar $r$-process abundance calculations and the average abundances from metal-poor halo stars are shown as a function of the solar $r$-process fraction, i.e., the fraction of $r$-process abundance in total neutron capture process. By comparing with the average abundances from metal-poor halo stars, it is found that the solar $r$-process abundances agree well with the stellar average abundances for the elements with large solar $r$-process fraction. Due to the large solar $s$-process fraction, the uncertainties of the solar $r$-process abundances are relatively large for elements Ba, Ce, La, Nd, Pr, and Hf, while the abundance consistency is still remained well for these nuclei within the uncertainties.

\begin{figure}[h]
  \includegraphics[width=8cm]{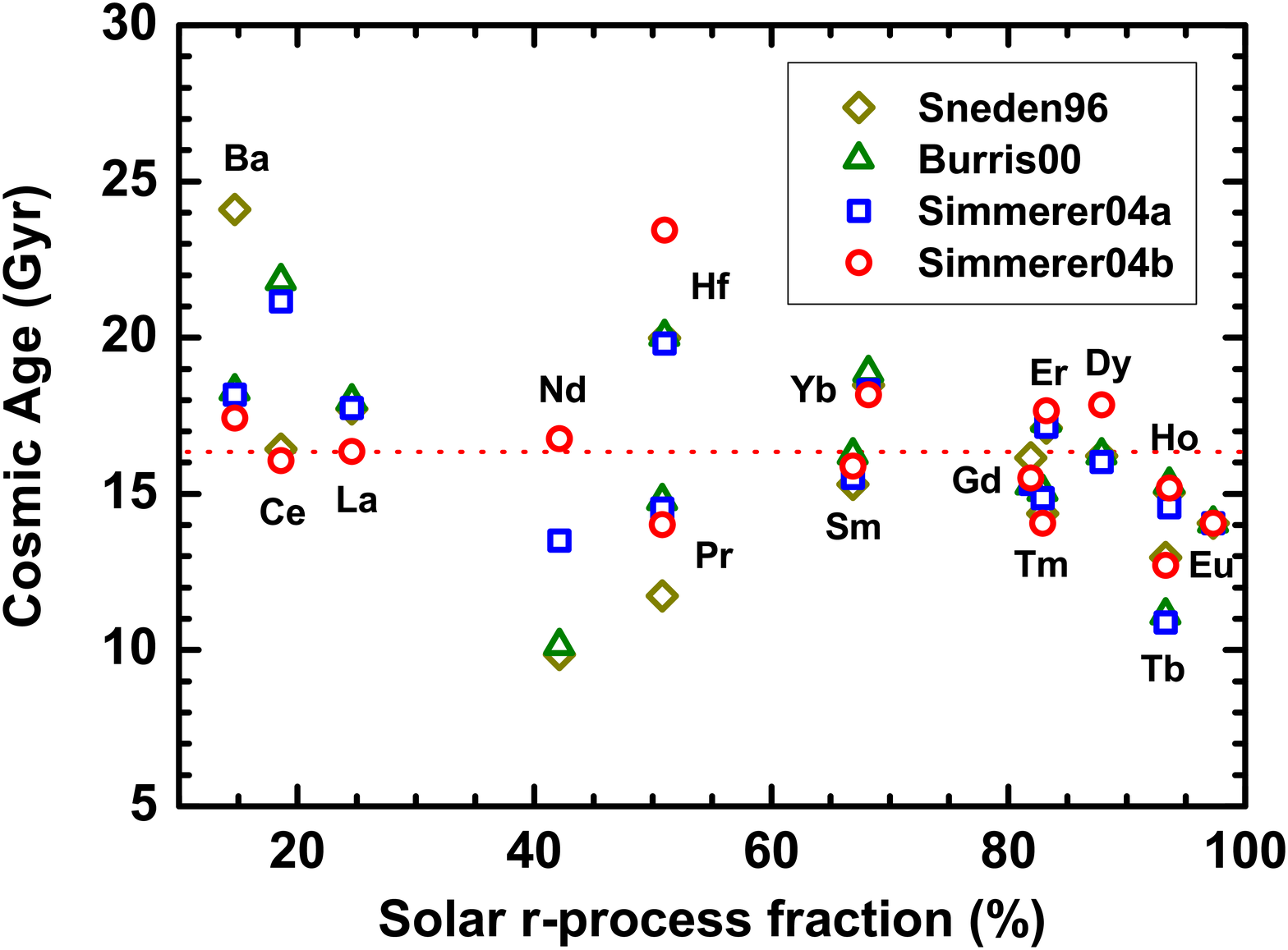}
  \caption{(Color online) The cosmic ages calculated by using average elemental abundances of five metal-poor halo stars and four data sets of solar $r$-process abundances. The dotted line denotes the average value of those ages shown with circles.}\label{fig2}
\end{figure}
Based on Eq.~(\ref{Eq:Age}), the ages can be estimated with various Th/X chronometers and the corresponding results are shown as a function of the solar $r$-process fraction in Fig.~\ref{fig2}. Clearly, the calculated ages are generally within $10-20$ Gyr for various chronometers. Considering the element whose solar $r$-process fraction exceeds $60\%$, the cosmic ages calculated by using the four sets of solar $r$-process abundances are relatively consistent. For those elements with $r$-process fraction less than $60\%$, deviations of cosmic ages obtained by utilizing different data sets can be clearly seen in Fig.~\ref{fig2}. This correlation between age uncertainties and the $r$-process fraction can be well understood by the abundance uncertainties shown in Fig.~\ref{fig1}. Therefore, those ages determined with the Th/X chronometers, whose fraction of $r$-process exceeds $60\%$ for the element X, are relatively reliable and we will take the ages determined with these Th/X chronometers as an extra group in the following.

\begin{figure}[h]
  \includegraphics[width=8cm]{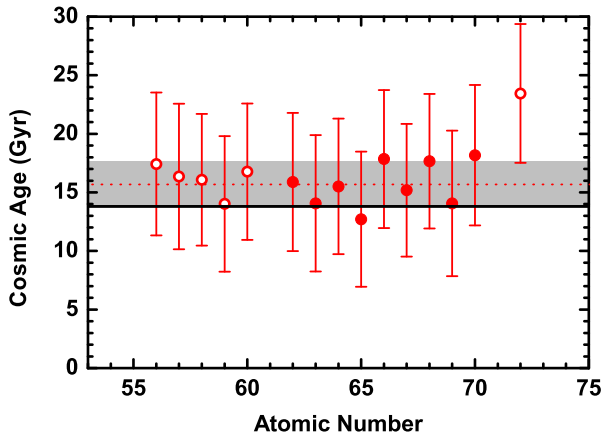}
  \caption{(Color online) The cosmic age calculated with different Th/X chronometers. For the element X whose $r$-process fraction exceeds 60\%, the corresponding Th/X age is denoted with the filled circle. Their average age and the corresponding error are denoted by the dotted line and the gray hatched area. For comparison, the cosmic age from Planck 2013 results is shown by the solid line.}\label{fig3}
\end{figure}
It can be clearly seen in Fig.~\ref{fig1} that the solar $r$-process abundances calculated on the
basis of stellar $s$-process model (solid line) can reproduce the stellar average abundances
(filled circles) well for the RE elements. Moreover, the $s$-process site has been well known to
be occurred in the AGB stars. Therefore, we take this solar $r$-process abundances as an example
in the following. The calculated cosmic ages are shown in Fig.~\ref{fig3} with the errors
estimated by using Eq.~(\ref{Eq:Err}). The filled circles denote the Th/X ages, whose $r$-process
faction exceeds $60\%$ for the elements X.

To estimate the stellar age more reliably, one could adopt the average value from various Th/X
chronometers, since the ages determined from different Th/X chronometers are consistent within
uncertainties. The corresponding uncertainty of the average value is calculated with $\sqrt{\sum_i
(\delta t)_i^2}/n$, where $(\delta t)_i$ denotes the error of the corresponding Th/X chronometer
and $n$ is the number of the Th/X chronometers used in the calculations. In this way, the average
age of all Th/X chronometers and the corresponding error are $16.35\pm 1.52$ Gyr. By taking the
Th/X ages, whose $r$-fraction exceeds $60\%$ for the element X, as a group, the average cosmic age
and the related error are $15.68\pm 1.95$ Gyr. This value is smaller than the average value on all
Th/X ages, but it is still larger than the latest cosmic age $13.813\pm 0.058$ Gyr determined from
Planck 2013 results~\cite{Planck2013arXiv}, while they are still consistent within the
uncertainties. It should be noted that if the cosmic age determined with the CMB data is indeed
smaller than the stellar age, then there would be something wrong about either the Big Bang theory
or the theory of radiometric dating. However, the uncertainty of $1.95$ Gyr is relatively large to
confirm this age deviation. Since this uncertainty of determined cosmic age mainly originates from
the error on thorium abundance observed in metal-poor star CS 22892-052, so future high-precision
abundance observations on CS 22892-052 are needed to understand this age discrepancy.

In summary, the lower limit of the cosmic age is estimated with radiometric method. By taking the
solar $r$-process abundances at the time when the Solar System became a closed system to
approximate the initial $r$-process abundances, a reliable age estimate for metal-poor halo star
is derived to be $15.68\pm 1.95$ Gyr without uncertainties from the theoretical $r$-process
calculations. This value is larger than the latest cosmic age $13.813\pm 0.058$ Gyr determined
from Planck 2013 results~\cite{Planck2013arXiv}, while they are still consistent within the
uncertainties. To confirm this age deviation, the uncertainty of $1.95$ Gyr should be further
reduced. Since this uncertainty mainly originates from the observed errors in the thorium
abundance of metal-poor star CS 22892-052, future high-precision abundance observations on CS
22892-052 are needed to understand this age deviation.

This work was partly supported by the National Natural Science Foundation of China under Grants
No. 11205004 and No. 11175001, the 211 Project of Anhui University under Grant No.
02303319-33190135.



\end{document}